\documentclass[prl,twocolumn,superscriptaddress,showpacs,amsmath,amssymb]{revtex4-2}
\usepackage[dvipsnames]{xcolor}
\usepackage{graphicx}
\usepackage{bm}
\usepackage{mathtools}
\usepackage{physics}
\usepackage{fixmath}
\usepackage[colorlinks=true,allcolors=blue]{hyperref}
\usepackage{siunitx}
\usepackage{textcomp,gensymb}

\definecolor{RO-color2}{named}{Green}
\definecolor{TTH-color2}{named}{Maroon}
\definecolor{MS-color2}{named}{Red}
\definecolor{SB-color2}{named}{Blue}

\newcommand{\Ms}[1]{M_{#1}}

\DeclareMathOperator{\diag}{diag}
\newcommand\commacirc{%
  \mathrel{\ooalign{\hss$,$\hss\cr%
  \kern0.0ex\raise0.6ex\hbox{\scalebox{0.7}{$\circ$}}}}}

\DeclareMathOperator{\sgn}{sgn}

\begin{document}
\title{Dynamics of two ferromagnetic insulators coupled by superconducting spin current}

\author{Risto Ojaj\"arvi}
\affiliation{Department of Physics and Nanoscience Center, University of Jyväskylä, P.O. Box 35 (YFL), FI-40014 University of Jyvaskyla, Finland}

\author{F.S.~Bergeret}
\affiliation{
Centro de F\'{i}sica de Materiales (CFM-MPC), Centro Mixto CSIC-UPV/EHU, Manuel de Lardizabal 5, E-20018 San Sebasti\'{a}n, Spain}
\affiliation{Donostia International Physics Center (DIPC), Manuel de Lardizabal 4, E-20018 San Sebasti\'{a}n, Spain}

 \author{M.A.~Silaev}
\affiliation{Department of Physics and Nanoscience Center, University of Jyväskylä, P.O. Box 35 (YFL), FI-40014 University of Jyvaskyla, Finland}
\affiliation{Moscow Institute of Physics and Technology, Dolgoprudny, 141700 Russia}
\affiliation{Institute for Physics of Microstructures, Russian Academy of Sciences, 603950 Nizhny Novgorod, GSP-105, Russia}

\author{Tero T. Heikkil\"a}
\affiliation{Department of Physics and Nanoscience Center, University of Jyväskylä, P.O. Box 35 (YFL), FI-40014 University of Jyvaskyla, Finland}

\date{\today}

\begin{abstract}
A conventional superconductor sandwiched between two ferromagnets can maintain coherent equilibrium spin current.
This spin supercurrent results from the rotation of odd-frequency spin correlations induced in the superconductor by the magnetic proximity effect. In the absence of intrinsic magnetization, the superconductor cannot maintain multiple rotations of the triplet component  but instead provides a Josephson type weak link for the spin supercurrent.  We determine the analogue of the current-phase relation in various circumstances and show how it can be accessed in experiments on dynamic magnetization. In particular, concentrating on the magnetic hysteresis and the ferromagnetic resonance response, we show how the spin supercurrent affects the nonequilibrium dynamics of magnetization which depends on a competition between spin supercurrent mediated static exchange contribution and a dynamic spin pumping contribution. Depending on the outcome of this competition, a mode crossing in the system can either be an avoided crossing or mode locking. 
\end{abstract}

\maketitle

Superconductivity is characterized by a $U(1)$ symmetry breaking complex order parameter and the dissipationless supercurrent proportional to its phase gradient. In the context of spin transport, analogous spin superfluidity has been discussed in various scenarios\,\cite{sonin2010-scs}.
Recent work on coherent spin transport in multilayers containing ferromagnetic (F) and superconducting (SC) elements has opened the question of dissipationless spin transport in superconductors \cite{jeon2018-esp,jeon2019-efe,jeon2020-tps}. While there are notions of dissipationless or conserved spin currents in such systems \cite{jacobsen2016-css,aikebaier2018supercurrent}, it has remained unclear in which sense such spin currents can be observed in experiments on dynamical magnetization. Here we clarify the situation by showing how spin supercurrents (SS) naturally arise in ferromagnetic resonance (FMR) experiments involving two or more magnets, how it is linked to the gradient of the direction of the triplet order parameter, and how it mediates magnetic interactions.  

What distinguishes the superconducting  currents from  normal persistent ones \cite{buttiker1983-jbs,levy1990-mmc,bleszynski2009-pcn} is their robustness against disorder and interactions and the large length scales on which they occur. In F/SC systems SS depends on the magnetic proximity effect and its range is set by the coherence length of the SC.

\begin{figure}
    \centering
    \includegraphics[width=\columnwidth]{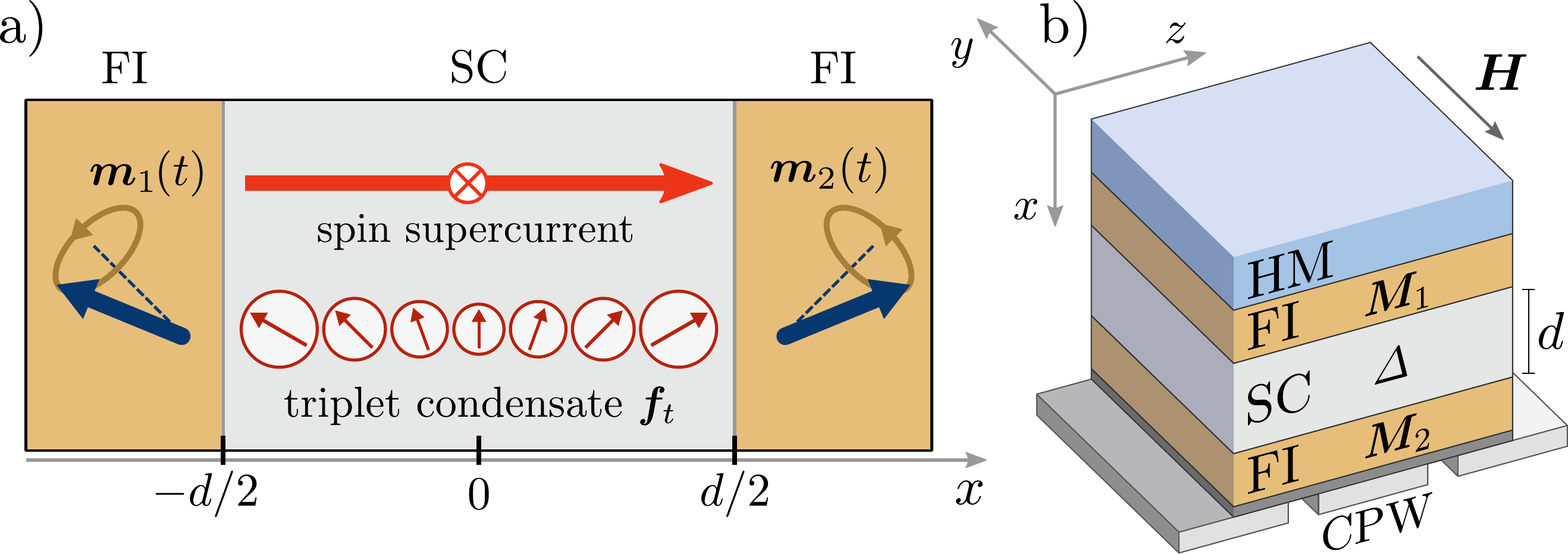}
    \caption{a) Spin supercurrent through a superconductor with thickness $d$. $\bm m_1(t)$ and $\bm m_2(t)$ are the instantaneous magnetization directions of the FIs, and the direction and radius of the `clocks` illustrate the spin direction and magnitude of the SC triplet condensate $\bm f_t$.
    b) Sketch of an FMR experiment. The sample is mounted on a coplanar waveguide (CPW). The additional heavy metal (HM) layer can be used to tune the Gilbert damping of the other FI.}
    \label{fig:spinsupercurrent}
\end{figure}

Here we study SS in possibly the simplest superconducting system in which it can exist, namely a SC sandwiched between two ferromagnetic insulators (FI) with noncollinear magnetizations (Fig.~\ref{fig:spinsupercurrent}). This system was considered already in the 1960s by de Gennes who showed that the SC mediates an antiferromagnetic interaction between the magnetic moments of the two FIs \cite{degennes1966coupling}. We demonstrate that this interaction can also be interpreted as an equilibrium spin current, and generalize it for a SC with a finite length and finite spin scattering. We consider the magnetization dynamics of two FIs coupled by spin pumping and SS, and show that spin supercurrents can lead to decreased or increased damping of the FMR modes of the trilayer, as compared to unhybridized modes in the normal (N) state. 

\begin{figure}
    \centering
    \includegraphics[width=\columnwidth]{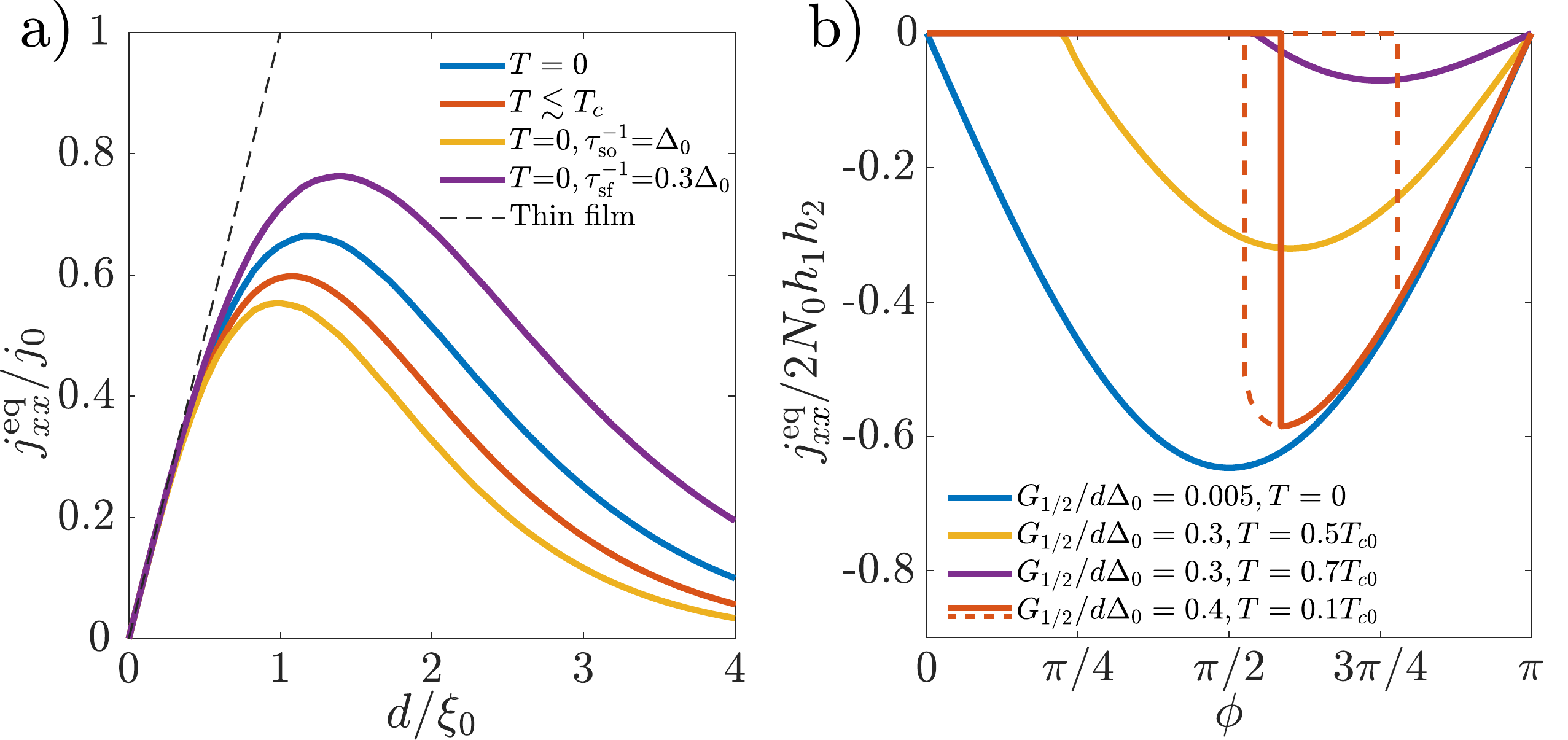}
    \caption{a) Spin supercurrent as a function of SC thickness with fixed $h_{n}\ll\Delta_0$
    normalized by $j_0 = h_1 h_2 \sin\phi \lim_{d\to 0}\left[ \chi(x_1,x_2) d\right] \xi_0$.
    b) Spin current--magnetization angle relation for $d=\xi_0$ with no spin scattering. Dashed lines indicate SC/N state hysteresis.}
    \label{fig:supercurrent_range}
\end{figure}

We first illustrate the concept of SS with a scheme based on the linearized Usadel equation. Consider a SC of length $d$ coupled with two non-collinear FIs, with magnetizations on $y$-$z$ plane pointing in directions $\bm m_{1}$ and $\bm m_2$ forming an angle $\phi$. The effective exchange field at the SC/FI interface leads to a partial conversion of the conventional singlet superconducting condensate into a triplet component \cite{bergeret2001long, bergeret2005-ots}. Thus, near $T_c$, the Matsubara Green's function (GF) for frequency $\omega_m>0$ has the general form
\begin{equation}
    \hat g = \tau_3 + (f_s + \bm f_t\cdot \bm\sigma)\tau_1,
\end{equation}
where $f_s$ and $\bm f_t$ are the singlet and triplet condensate functions, respectively.\cite{linder2019odd}
We assume 
translational invariance on the $y$-$z$ plane. The spin supercurrent in $x$-direction can then be expressed as \cite{ouassou2019-css}
\begin{equation}
\begin{split}
\bm j_{x}^{\rm eq} &= \pi T \sum_{\omega_m>0} N_0 D \bm f_t \times \partial_x \bm f_t\\
&= \pi T\sum_{\omega_m>0} N_0  D |\bm f_t|^2 \partial_x \varphi \,\hat{\bm x},
\end{split}\label{eq:spin_supercurrentTc}
\end{equation}  
where $\varphi$ is the angle of $\bm f_t$ relative to the $z$-axis and $N_0$ is the density of states at the Fermi level in the normal state. The SS
arises from the coherent rotation of the triplet Cooper pairs. The vector structure of $\bm j_x$ indicates its spin direction, and the subscript refers to the spatial direction.

The triplet condensate is determined by the Usadel equation and its boundary condition (BC)
\begin{subequations}
\begin{align}
    D \nabla^2 \bm f_t &= 2\omega_m \bm f_t,\label{eq:usadeltTc}\\
    \left.D\,\bm n\cdot\nabla \bm f_t\right|_{x{=}x_{n}} &= 2i G_{n} \bm m_{n} f_s,\label{eq:usadelbctTc}
\end{align}
\end{subequations}
where $D$ is the diffusion constant, $\bm n$ is the interface normal from FI to SC, and $x_{1,2} = \mp d/2$. The interface parameters $G_{n}$, which are related to the imaginary spin mixing conductances of the interfaces by ${\rm Im}\,G^{\uparrow\downarrow}_n = G_{n}N_0$, determine the coupling between the singlet and triplet Cooper pairs\,\cite{tokuyasu1988-pef,huertas2002absolute}.
To first order in $G_{n}$, the singlet retains its bulk value $f_s = \Delta/{\omega_m}$.
The triplet condensate leads to  a spin density \cite{bergeret2004-ifs,xia2009-ipe}
\begin{equation}
\begin{split}
\label{eq:spin_density}
    \bm S(x) &= 2i N_0 \pi T \sum_{\omega_m>0} f_s \bm f_t(x)\\
    &= \sum_{n=1,2} \chi(x,x_n) G_n \bm m_n.
\end{split}
\end{equation}
On the second line, we define the nonlocal spin susceptibility $\chi(x,x')$.

Substituting Eq.~\eqref{eq:usadelbctTc} into Eq.~\eqref{eq:spin_supercurrentTc}, we find the SS
\begin{equation}
    j_{xx}^{\rm eq} = G_1 G_2 \chi(x_1,x_2)  \sin\phi.\label{eq:supercurrentrangeTc}
\end{equation}
In analogy to SNS-junction, the FSF trilayer can be considered a spin Josephson junction\,\cite{fomin1991-scs,nogueira2004-sje}. The middle layer is weak in the sense that it cannot support multiple phase windings, since there is no energy penalty for the vanishing condensate function. In order to have strong SS with a total phase winding larger than $\pi$ would require the condensate to be restricted to $U(1)$ symmetry, as happens e.g. in easy-plane ferromagnets\cite{sonin2010-scs,takei2014-sst}. Figure \ref{fig:supercurrent_range}b shows the current-angle relation, which deviates from sine function at high $T$ and at strong coupling.

As an equilibrium current, $\bm j^{\rm eq}_{x}$ is conserved through the SC.\cite{ouassou2017-tcp} Figure \ref{fig:supercurrent_range}a shows its magnitude as a function of $d$. In the thin-film limit $d\ll\xi_0$, where $\xi_0$ is the SC coherence length, each interface induces a homogeneous exchange field with amplitude $h_{n}=G_{n}/d$.\,\cite{heikkila2019-tes} The total exchange field is $\bm h = h_1\bm m_1 + h_2\bm m_2$ is limited by the Chandrasekhar-Clogston critical value $h_c=\Delta_0/\sqrt{2}$. With fixed $h_{n}$ the SS is proportional to $d$.
When $d\to\infty$, the supercurrent vanishes as $\exp(-d/\xi_0)$. This exponential factor measures the overlap between the two proximity fields. Maximal SS is obtained when the maximum volume of SC is proximitized by both FIs at $d\approx \xi_0$.

At arbitrary temperatures and with finite spin-orbital/spin-flip scattering times $\tau_{\rm so/sf}$, 
we determine the SC spin accumulation in the dirty limit from the full Usadel equation\,\cite{supplementary}, using the spin-mixing 
BC
\cite{tokuyasu1988-pef,silaev2020-scc}
 \begin{align} \label{eq:usadel_BC}
  \left.D\,\bm n\cdot\check g \circ \nabla \check g\right|_{x=x_{n}} =
  iG_{n}
  [\bm m_{n}\cdot{\bm \sigma}\tau_3\commacirc \check g  ]  
  \,,
  \end{align}
where $\check g$ are the $8\times8$ quasiclassical Green's functions in Keldysh-Nambu-Spin space and $\circ$-products are time-convolutions. The above BC gives Eq.~\eqref{eq:usadelbctTc} as a special case. In the BC we take into account only 
the 
effective exchange field, 
\cite{tokuyasu1988-pef} and neglect terms associated with interfacial spin relaxation and decoherence.  \cite{bergeret2012electronic,eschrig2015general,zhang2019theory}

By taking the trace over spin of \eqref{eq:usadel_BC}, and integrating over the energy, the spin current through the $n$th interface is  \cite{ojajarvi2021-ges}
\begin{equation}
\bm j_{x,n}(t) = G_{n} \left[ \bm S(t,x_n) \times \bm m_n(t) - N_0\dot{\bm m}_n(t)\right],\label{eq:boundary_current}
\end{equation}
where  $\bm S$ contains both the equilibrium spin density 
{\it cf.} Eq. (\ref{eq:spin_density})
and the non-equilibrium spin accumulation. We assume that $d\gg\lambda_F$, where $\lambda_F$ is Fermi wavelength, and neglect the short range Pauli paramagnetic contribution. The $\dot{\bm m}_n$ term is the spin pumping contribution, and $\bm S\times\bm m_n$ gives the equilibrium SS and the back-action due to non-equilibrium spin accumulation induced by spin pumping. \cite{tserkovnyak2005-nmd} According to Eq.~\eqref{eq:boundary_current} there is a finite SS if the equilibrium $\bm S$ and the magnetization of the FI are not collinear.

We introduce a Stoner-Wohlfarth\,\cite{stoner1948-mmh} type free energy (per interface area) to describe the effect of SS on the magnetic configuration of the FI/SC/FI trilayer
\begin{align}
F(\bm m_1, \bm m_2) &= \sum_{n=1,2} \Big[ -\mu_0{\bm M_{n}{\cdot}\bm H} -  K_A^{n}(\bm m_n \cdot \bm z)^2 \nonumber\\
&\mkern-20mu+ {K_B^n(\bm m_n\cdot\bm x)^2} \Big] + F_{\rm sc}(\bm m_1\,{\cdot}\,\bm m_2, \Delta),\label{eq:sw_free_energy}
\end{align}
where $|\bm M_n|=M_n$ 
 is the FI magnetic moment per interface area. 
The free energy includes the Zeeman energy from external magnetic field $\bm H$, the SC free energy $F_{\rm sc}$, and the in-plane easy axis/out-of-plane anisotropy energies $K_{A/B}^n$. 
In the thin-film limit, at $T=0$ and without any spin scattering $F_{\rm sc}(\bm h) = d N_0 (|\bm h|^2 - \Delta^2/2)$. In the general case we use the SC energy functional of Ref.~\onlinecite{virtanen2020-qfe}.

In a static setting, the coupling between the magnets can be described as an effective magnetic field $\mu_0\bm H_{\rm eff,1}^{\rm sc} = 
-J_\phi\bm m_{2}/\Ms{1}$, which can be identified with SS via the spin-transfer torque
\begin{equation}
\begin{split}
\bm j_{x,1}^{\rm eq} &= -\bm M_1 \times \mu_0\bm H_{{\rm eff},1}^{\rm sc}\\
&= J_\phi\,\bm m_1\times\bm m_2,\label{eq:stt}
\end{split}
\end{equation}
with exchange constant $J_\phi={\rm d}{F_{\rm sc}}/{\rm d}{\cos\phi}\geq 0$. At weak coupling Eq.~\eqref{eq:stt} coincides with Eq.~\eqref{eq:supercurrentrangeTc}.

\begin{figure}
    \centering
    \includegraphics[width=\columnwidth]{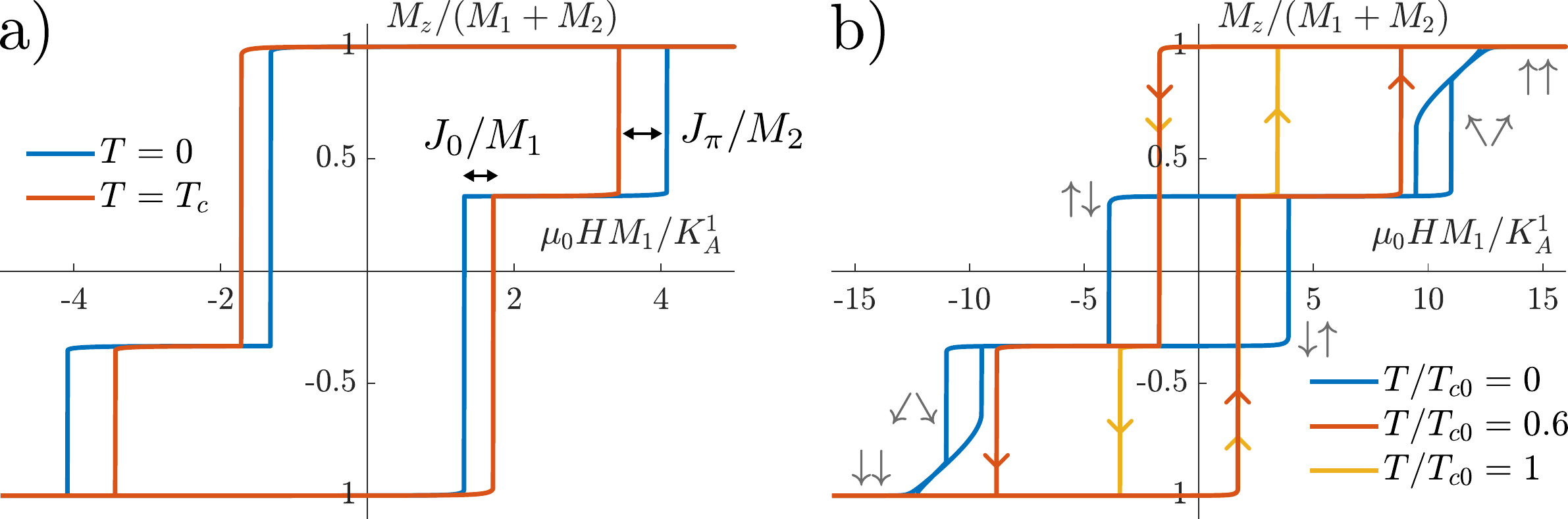}
    \caption{Hysteresis loops for two coupled ferromagnets. a) Weak SC exchange interaction $d N_0\Delta_0^2/K_A^n=1$. For $T=0$ the system is always in the SC state. b) Strong exchange interaction $d N_0\Delta_0^2/K_A^n=15$. For $T/T_c=0.6$ the system is in the N state in P configuration, but in the SC state for AP configuration. 
    For both panels, $M_1=2M_2$, 
    $h_n=0.3\Delta_0$, $\tau^{-1}_{\rm so/sf}=0$. External field is at {2\degree} angle to the anisotropy axis.
    }
    \label{fig:hysteresis}
\end{figure}

When the SC exchange energy is small compared to the  anisotropy energies, SS modifies the coercive fields \cite{zhu2017-sec} (Fig.~\ref{fig:hysteresis}a).
In the SC state, the coercive fields for AP-to-P switching increase by $J_\pi/\Ms{n}$, and the coercive fields for P-to-AP switching decrease by $J_0/\Ms{n}$ relative to the N state. 
With a strong SS (Fig.~\ref{fig:hysteresis}b) the anisotropies cannot force the magnets into a binary parallel/antiparallel configuration space and the exchange interaction may induce a spin-flop transition, in which the two magnets collectively 
rotate from AP to P configuration\, \cite{wigen1993-dra}.

At low temperatures the SC transition is of the first order as a function of $\phi$ and exhibits hysteresis (Fig.~\ref{fig:supercurrent_range}b).\cite{maki1964-ppm,tedrow1970experimental,wu1995asr,butko1999state} If there is a strong uniaxial anisotropy in the FIs, 
it is possible to study the N-to-SC hysteresis\cite{wu1995asr,butko1999state}, as opposed to the ordinary magnetic hysteresis resulting from anisotropies. By applying the external magnetic field perpendicular to the easy axis direction, the exchange field changes continuously. If the SC transition is continuous, there is also no magnetic hysteresis. If the transition is of the first order, the N-to-SC and magnetic hysteresis become entangled. We show in the supplementary how the two can be disentangled in the magnetization curve.\cite{supplementary}


We now consider the dynamical effects due to SS. In contrast to most studies of F/SC hybrid structures \cite{simensen2021-spb,montiel2021-sci}, we take both magnets as dynamical. We study the small-angle dynamics around the 
equilibrium configuration with the ansatz
\begin{equation}
    \bm m_n(t) = \bm m_n(0) + \Re[\bm m_n(\omega) \exp(-i\omega t)],
\end{equation}
where $\bm m_n(\omega)$ is a small perturbation perpendicular to $\bm m_n(0)$.
The dynamical magnetization induces a time-dependent exchange field 
at the SC interfaces, which in turn induces a time-dependent spin density in the SC.

The dynamics of the FI magnetizations are described by the classical Landau-Lifshitz-Gilbert (LLG) equation, supplemented by a spin current term \cite{slonczewski1996-cde}
\begin{equation}
    \dot{\bm M}_n = -\gamma\bm M_n \times \mu_0\bm H_{\rm eff,n}^{(0)}
    + \frac{\alpha_n}{\Ms{n}} (\bm M_n\times \dot{\bm M}_n) + \gamma \bm j_{x,n},\label{eq:LLG}
\end{equation}
where $\bm H_{\rm eff,n}^{(0)}$ is the effective magnetic field in the N state including the external magnetic field and anisotropy fields, $\gamma$ is the gyromagnetic ratio, and $\alpha_n$ is the 
Gilbert damping coefficient, which can be controlled with an additional heavy metal layer next to the FI (Fig.~\ref{fig:spinsupercurrent}b). The spin current 
is given by Eq.~\eqref{eq:boundary_current}.

In a FMR experiment, the external field has a static 
component and a small transverse dynamic part $\bm H_{\rm rf}$. The linear response to $\bm H_{\rm rf}$ 
given by LLG Eq.~\eqref{eq:LLG} is
\begin{equation}
    \hat\chi^{-1}_n(\omega)\bm M_n(\omega) = \mu_0 \bm H_{\rm rf}(\omega) - \bm m_n(0)\times\frac{\bm j_{x,n}(\omega)}{\Ms{n}},\label{eq:magn_circ}
\end{equation}
where $\hat \chi^{-1}_{n}$ are the $2\times2$ magnetic susceptibilities of the uncoupled magnets.\,\cite{supplementary} Since the magnets are insulating, there are no eddy currents inside the FIs and we can neglect the direct coupling between the SC and the rf field.\cite{muller2021temperature,kennewell2007calculation}

In the parallel configuration, the coupled dynamics of the two magnets 
is described by a 
matrix susceptibility
 \begin{equation}
     \hat\chi^{-1}_{\rm tot}(\omega) = \begin{pmatrix}
     \hat\chi^{-1}_1(\omega)+\frac{J_0 - \delta \hat J_{11}(\omega)}{\Ms{1}^2} & -\frac{J_0 + \delta\hat J_{12}(\omega)}{\Ms{1}\Ms{2}}\\
     -\frac{J_0 + \delta \hat J_{21}(\omega)}{\Ms{1}\Ms{2}} & \hat\chi^{-1}_2(\omega)+\frac{J_0-\delta \hat J_{22}(\omega)}{\Ms{2}^2}
     \end{pmatrix},\label{eq:magnetic_susceptibility}
 \end{equation}
where $J_0=-G_1 G_2 \chi(x_1,x_2)$ is the static exchange constant, and
\begin{equation}
\begin{split}
\delta \hat J_{ij}(\omega) &= -G_i G_j\left[ \hat\chi(\omega,x_i,x_j)\,{-}\,\chi(x_i,x_j)\right]\\
&- N_0 G_i \omega \delta_{ij} \hat\sigma_3,\label{eq:dynamic_coupling}
\end{split}
\end{equation}
are the dynamic corrections related to spin pumping and other finite-frequency processes. Here, $\hat\chi$ is the dynamic spin susceptibility,\,\cite{silaev2020-ffs,supplementary} related to the static spin susceptibility by $\hat \chi(0,x_i,x_j)= \chi(x_i,x_j)\hat 1$.

\begin{figure}
    \centering
    \includegraphics[width=\columnwidth]{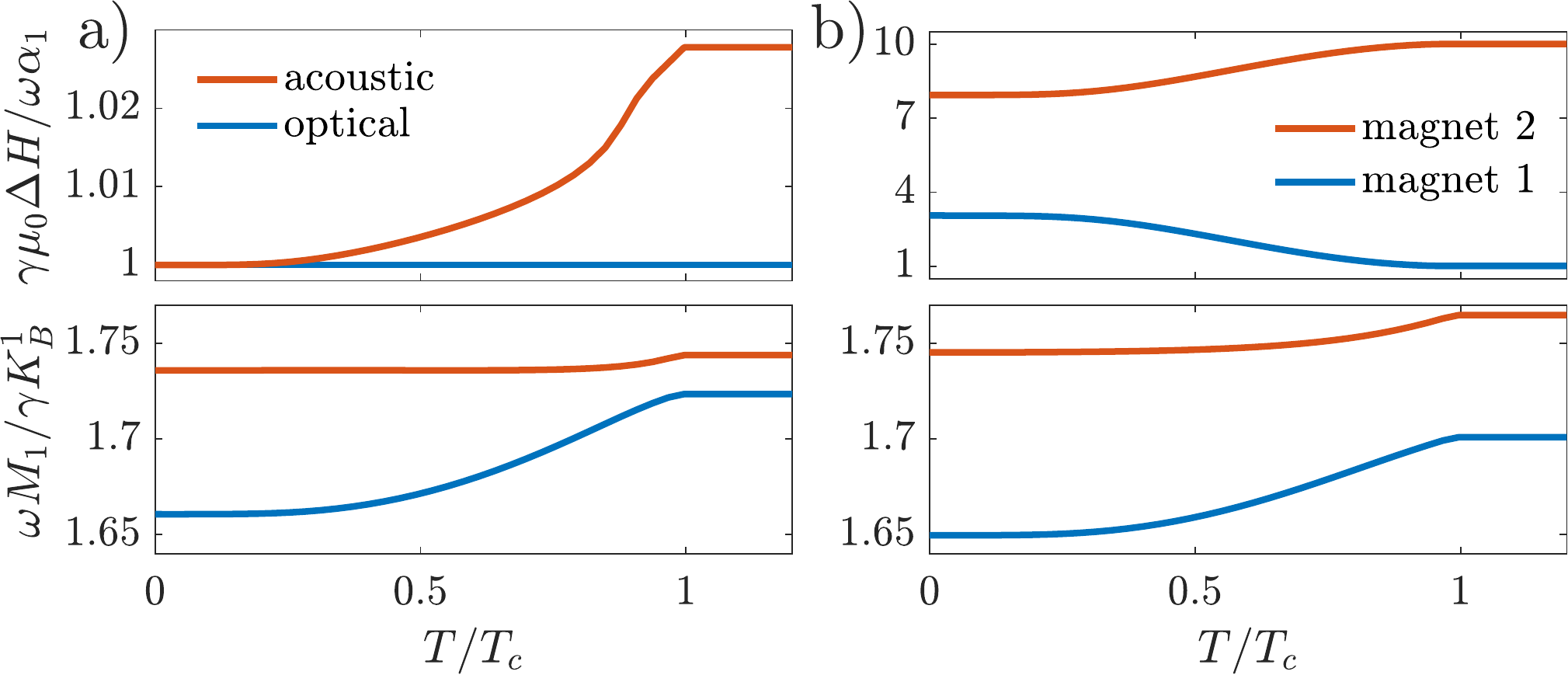}
    \caption{Temperature dependence of linewidth (top) and resonance frequency (bottom) for a) symmetric system with $\alpha_{n}=0.005$,
    b) asymmetric system with $\alpha_1=0.005$ and $\alpha_2 = 0.05$. Parameters: $K^{n}_A=0$, $d N_0\Delta_0^2=0.1$, $\Delta_0 M_n/\gamma=100$, $\gamma\mu_0 M_{n} H=1$, in units of $K_B^{n}$; $h_{n}=0.3\Delta_0$, $\tau_{\rm so}^{-1}=0.2\Delta_0$}
    \label{fig:damping}
\end{figure}

To illustrate the effect of SS on the FMR properties, we first consider a fully symmetric trilayer. In the P configuration, the eigenmodes are the acoustic and optical modes for which $\bm m_1(t) = \pm\bm m_2(t)$, respectively.\,\cite{wigen1993-dra} In the acoustic mode, the magnetizations are always parallel to each other and there is no SS\,\cite{wigen1993-dra}. The magnets are only coupled by the residual part of the susceptibility, $\chi(\omega)-\chi(0)\approx \omega \chi'(0)$. The imaginary part $\Im\chi'(0)$ contributes directly to dissipation and can be included in the Gilbert damping coefficient. It describes the relaxation of quasiparticles, and vanishes at low $T$ where quasiparticles cannot be excited due to SC gap\,\cite{morten2008-pea}. This leads to the usual decrease of the FMR linewidth in the SC state (Fig.~\ref{fig:damping}a).
The real part $\Re\chi'(0)$ shifts the resonance frequency. 

In the optical mode the magnetizations precess out-of-phase and are strongly coupled by the SS. In this case, the effective magnetic field is shifted by $2 J_0/\Ms{n}$. The resonance field difference between acoustic and optical modes at fixed frequency gives a direct measure of SS. However, measuring the optical mode can be difficult as a symmetrically applied rf field excites only the acoustic mode. Optical mode can be excited by longitudinal FMR pumping\,\cite{zhang1994-ufr,lindner2003situ}, or by breaking the symmetry. For the optical mode the non-equilibrium spin currents pumped by the two magnets partially cancel in the SC spacer. In the thin-film limit this cancellation is exact and the  dissipation in the spacer does not affect the linewidth.

In an asymmetric trilayer, SS can have a drastic effect on the linewidths of the FMR modes. For illustration, let us neglect the spin pumping and consider only the effect of SS together with the intrinsic damping of the magnets. In the N state the magnets are uncoupled, and the eigenmodes are the Kittel modes of the individual magnets with linewidths $\Delta H_{n}^0$ proportional to Gilbert damping constants $\alpha_n$. In the SC state, the SS hybridizes the modes so that their linewidths become weighted averages $\Delta H_{n} = (\Delta H_1^0 + p_{n} \Delta H_2^0)/(1+p_{n})$, where $p_{1} = 0$ and $p_{2} = \infty$ in the uncoupled system, and $p_{1} \to \Ms{1}/\Ms{2}$ and $p_{2} \to \Ms{2}/\Ms{1}$ in the strongly coupled system.\cite{layadi2015-tif} The top panel of Fig.~\ref{fig:damping}b shows a numerical evaluation for the linewidth of such a system, including the spin pumping contribution. In particular, if one magnet has a lower intrinsic damping than the other magnet, the linewidth of the related mode increases below the SC transition.

\begin{figure}
    \centering
    \includegraphics[width=\columnwidth]{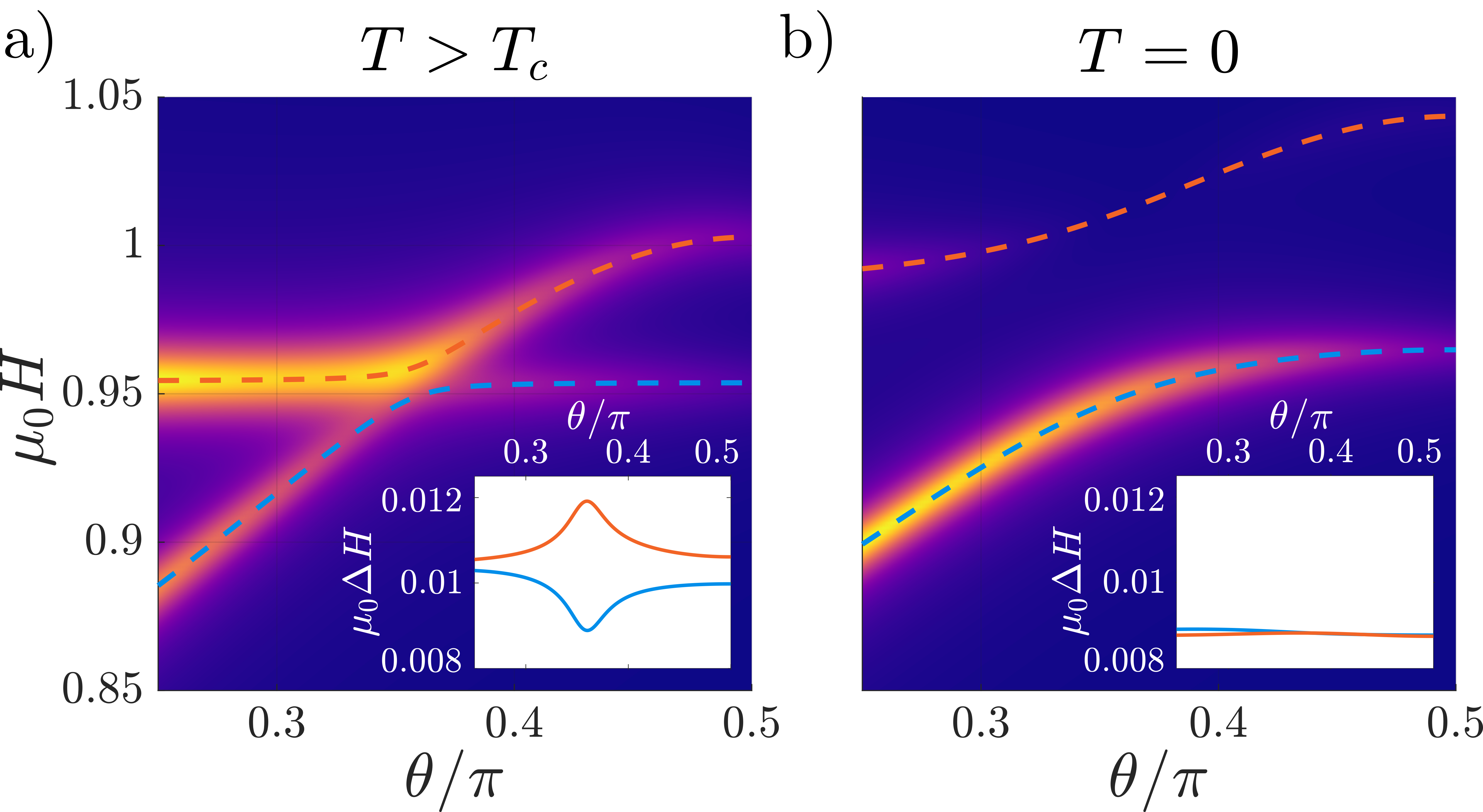}
    \caption{
    Absorbed rf power as a function of external field angle $\theta$ and amplitude $\mu_0 H$ (in units of $K_B^1/M_1$). Dashed lines indicate the resonance fields and the insets show the linewidths. 
    a) Mode locking in the N state is the most evident from the linewidth broadening/narrowing near the crossing. 
    b) Avoided crossing in the SC state. The linewidths do not depend strongly on the proximity to the crossing. Parameters are as in Fig.~\ref{fig:damping}a except for
    $\omega M_1/\gamma=1.75$,
    $K^{1}_A = 0.1$, 
    $K^2_B=1.1$, 
    $\Delta_0 M_n/\gamma=10$, given in units of $K_B^1$.
    }
    \label{fig:crossing}
\end{figure}

The SS-mediated exchange interaction and spin pumping depend on temperature in opposite ways; spin pumping vanishes at zero temperature, whereas SS vanishes in the normal state. The competition between these two processes can be studied at a mode crossing between two FMR modes (Fig.~\ref{fig:crossing}), which can be engineered e.g. by having FI films with different thicknesses. In general, thinner films will have stronger anisotropy fields. A mode crossing may then be seen by rotating the applied in-plane magnetic field relative to the anisotropy axis. \cite{heinrich2003dynamic}

In the N state (Fig.~\ref{fig:crossing}a) the magnets are only coupled by spin pumping. The dissipative component of spin pumping generated by spin relaxation in the SC layer gives rise to mode attraction. Its signature is the sudden change of the mode linewidths at the mode crossing (inset of Fig~\ref{fig:crossing}a) \cite{tserkovnyak2005-nmd,tserkovnyak2020-epd}. 
In the SC state (Fig.~\ref{fig:crossing}b), the SS mediated exchange coupling dominates over spin pumping, creating a regular avoided crossing.

So far the dynamical properties of SS have been experimentally studied only in systems with ferromagnetic metals (FM)\,\cite{jeon2018-esp,jeon2019-efe,jeon2020-tps}, and an increase in FMR linewidth below $T_c$ has been observed in systems which include multiple FM or heavy metal layers with strong spin-orbit coupling. In some of these experiments\cite{jeon2018-esp}, there is nominally only a single magnet, but the heavy metal layers close to the ferromagnetic instability can be magnetized by the SS-mediated magnetic proximity effect.\cite{montiel2018-gps} 

With minor changes, our theory can describe SS-mediated coupling in FM/SC/FM trilayer. The form of the magnetic susceptibility \eqref{eq:magnetic_susceptibility} is otherwise unchanged from the insulating case, but the susceptibilities for the spin density at the interfaces are replaced by the susceptibilities for the total spin density of the FM conduction electrons, and the spin mixing conductance is replaced by the $s$-$d$ coupling inside the FM.
In contrast to an FI, in FM the spin current is not absorbed in a layer of atomic thickness, but penetrates into the FM at the range of $\xi_F=\sqrt{D_F/2T}$, where $D_F$ is the diffusion coefficient of the FM. \cite{bergeret2005-ots} The long-range triplet component\cite{bergeret2005-ots} penetrating deep into the FM is exactly the component non-collinear to its exchange field, and the one related to SS. Because FMs support eddy currents, there is also an electromagnetic coupling between the layers.\cite{muller2021temperature,kennewell2007calculation}. Finding the magnitude of spin currents in a metallic system will be left for further work. 

Despite these differences, our framework suggests that the experimentally observed enhancement below $T_c$ is likely to be a result of SS-mediated hybridization between the FMR modes. In interpreting the FMR data in systems with superconducting interlayers and multiple magnets, one should not rely on the spin pumping picture with a single dynamical magnet, but instead model the magnetization dynamics of the whole structure.

{\it Conclusions}. We have studied the properties of SS in FI/SC/FI systems and shown how they can mediate damping between ferromagnetic insulators even though they, as equilibrium currents, are themselves non-dissipative. In an analogy to Josephson junctions, the SS can be characterized via 
the spin current - magnetization angle relation. This can be accessed by studying temperature dependent modifications to the FMR 
frequencies in FI/SC/FI setups. 

The phenomena described here can be studied in various different FI/SC combinations, provided they can be suitably stacked or placed next to each other. The FI can be for example GdN,  EuS/EuO, or  ferrimagnetic YIG,  which have been recently studied in combination with superconductors such as Nb, NbN or Al. \cite{yao2018-psd,hijano2021-csc,zhu2017-sec,li2013-sss,cascales2019switchable,strambini2017revealing,rogdakis2019-stp} Our results can be used to design magnetic resonator structures where the SS mediates a tunable coupling between the resonators.

\begin{acknowledgments}
      {\bf Acknowledgements}
  This work was supported by 
  the Academy of Finland Project 317118, 
  the European Union’s Horizon 2020 Research and Innovation Framework Programme under Grant No. 800923 (SUPERTED), and 
  Jenny and Antti Wihuri Foundation. FSB acknowledges funding by the Spanish Ministerio de Ciencia, Innovacion y Universidades (MICINN) (Project FIS2017-82804-P).
\end{acknowledgments}

%

\clearpage
\appendix

\section{Supplementary material}

\renewcommand{\theequation}{S\arabic{equation}}
\setcounter{equation}{0}
\renewcommand{\thefigure}{S\arabic{figure}}    
\setcounter{figure}{0}    

The FI/SC/FI trilayer is described by the Hamiltonian
\begin{equation}
    H = H_{\rm SC} + H_{\rm imp} + \sum_{n=1,2} \left(H_{\rm FI}^{(n)} + H_{\rm ex}^{(n)}\right),
\end{equation}
where $H_{\rm SC}$ is the BCS Hamiltonian for an $s$-wave superconductor, assumed infinite in $y$ and $z$ directions, but with a finite thickness $d$ in the $x$ direction.
$H_{\rm imp}$ is the random impurity potential in the SC, containing both non-magnetic and magnetic impurities, which gives the spin-orbit and spin-flip self-energies, respectively. We assume that the elastic mean-free path is short so that the SC can be described with the quasiclassical Usadel equation.

The Hamiltonians describing the FIs and the SC/FI interfaces are\cite{heikkila2019-tes}
\begin{align}
    H_{FI}^{(n)} &= -\sum_{i,j} J_{ij} \bm S_{\bm i}\cdot \bm S_{\bm j} + H_{\rm an},\\
    H^{(n)}_{\rm ex} &= - J_{{\rm ex},n} \sum_{i\in\mathcal I_n} \Psi^\dag(\bm r_i) \bm S_{i}\cdot \bm \sigma \Psi(\bm r_i),
\end{align}
where $J_{ij}$ is the exchange coupling between the localized spins $\bm S_{i}$ at FI lattice sites $\bm r_i$. $H_{\rm an}$ includes the possible anisotropy fields.
$\mathcal I_n$ is the set of spins at the interface between $n$th FI and the SC, and the spin-mixing conductances are related to the interfacial exchange coupling $J_{{\rm ex},n}$ by $G_n = J_{{\rm ex},n} S_n a$, where $S_n$ is the local spin at the interface and $a$ is the FI lattice constant.\cite{heikkila2019-tes}
 The Nambu field operator
\begin{equation}
    \Psi^\dag(\bm r) = \begin{pmatrix}
    \psi^\dag_\uparrow(\bm r) & \psi^\dag_\downarrow(\bm r) & -\psi_\downarrow(\bm r) & \psi_\uparrow(\bm r)
    \end{pmatrix}
\end{equation}
is chosen so that the $s$-wave SC order parameter is proportional to unit matrix in spin space. 

\subsection{Quasiclassical theory}

When the elastic mean-free path is short, the SC specified by the above Hamiltonian can be described at the quasiclassical limit by using the Keldysh-Usadel equation\,\cite{usadel1971-mds, heikkila2019-tes,ojajarvi2021-ges} 
  \begin{equation}\label{eqS:KeldyshUsadelFI}
  - \{\tau_3\partial_t\commacirc \check g \} + 
  \partial_x( D\check g\circ \partial_x \check g)  = 
   [ \Delta \tau_1 + \check \Sigma\commacirc \check g\, ]
  \end{equation} 
with the boundary conditions given by
Eq.~(6) of the main text.
Above, $\check g$ is the quasiclassical Green's function (GF) in 8$\times$8 space consisting of Keldysh, Nambu and spin indices. It obeys normalization condition ${\check g\circ \check g} = \delta(t,t')$.
 The elastic spin relaxation is determined by spin-orbit and spin-flip scattering self-energies $
    \check \Sigma = \hat{\bm\sigma}\cdot
  \check g \hat{\bm\sigma}/{6\tau_{\rm so}} + \tau_3\hat{\bm\sigma}\cdot
  \check g \hat{\bm\sigma}\tau_3/{6\tau_{\rm sf}},
 $ where $\tau_{\rm so/sf}$ are the scattering rates. The order parameter $\Delta$ is solved from the self-consistency equation
\begin{equation}
\label{eqS:selfconsistency}
    \Delta(\omega) = \frac{\lambda}{16i}\int_{-\Omega_{\rm  D}}^{\Omega_{\rm  D}} d\varepsilon \Tr[\hat\tau_1 \hat g^K(\varepsilon,\varepsilon-\omega)],
\end{equation}
where the coupling constant $\lambda$ and the cutoff $\Omega_{\rm D}$ can be eliminated in favor of SC order parameter $\Delta_0$ at $T=0$ in the absence of pair-breaking effects.\,\cite{heikkila2019-tes}.

The spin current density in $x$-direction and the spin density are defined as
\begin{align}
    \bm j_{x}(\omega,x) &= \frac{N_0}{16}\int_{-\infty}^\infty d\varepsilon \Tr[\bm\sigma D (\check g\circ \partial_x \check g)(\varepsilon,\varepsilon{-}\omega)]^K,\\
    \bm S(\omega,x) &= -\frac{N_0}{8}\int_{-\infty}^\infty d\varepsilon \Tr[\bm\sigma\tau_3 \hat g^K(\varepsilon,\varepsilon{-}\omega;x)].
\end{align}

For the SC we use the free energy functional\,\cite{virtanen2020-qfe,aikebaier2019-snm}
\begin{align}
    \label{Eq:FreeEnergyGen0}
  &F_{\rm sc}[\hat g,\Delta] =  N_0\int{\rm d}x \Big(\frac{\Delta^2}{\lambda} \\ 
  & -\frac{\pi T}{2} \sum_{\omega }
     {\rm tr}
     \{ (\omega_n + i \bm{h}\cdot\bm{\sigma})\hat{\tau}_3 
     \hat g
     + \Delta\tau_1\hat{g}  -\frac{D}{4} (\hat{\nabla} \hat g)^2 \nonumber\\
     &+ \frac{1}{12\tau_{\rm so}}(\bm{\sigma}\hat{g})\cdot(\bm{\sigma}\hat{g}) + \frac{1}{12\tau_{\rm sf}}(\bm{\sigma}\hat{\tau}_3\hat{g})\cdot(\bm{\sigma}\hat{\tau}_3\hat{g})
  \}\Big),\nonumber
\end{align}
which is regularized by subtracting the normal state energy with $\hat g =  \sgn(\omega_n)\hat\tau_3$ and $\Delta=0$.

\subsection{Spin susceptibility at the thin-film limit}

In equilibrium, spatially averaging over Eq.~\eqref{eqS:KeldyshUsadelFI} 
 and using the BC 
 of Eq.~(6),
 we find an effective position-independent Usadel equation
  \begin{equation}\label{eqS:KeldyshUsadelFI2}
  [ i(\varepsilon - \bm h\cdot\bm\sigma)\tau_3 + \Delta \tau_1 + \hat \Sigma, \hat g\, ] =0.
  \end{equation} 
This equation is valid at the thin-film limit $d\ll\xi_0,l_{\rm sd}$, where $l_{\rm sd}=[D/(\tau_{\rm so}^{-1}+\tau_{\rm sf}^{-1})]^{n}$ is the normal state spin diffusion length. Equations \eqref{eqS:selfconsistency} and \eqref{eqS:KeldyshUsadelFI2} together with the normalization condition constitute a nonlinear group of equations which we solve numerically. In the thin-film limit we drop the spatial indices and define the static susceptibility as
$\bm S = \hat\chi(0) \bm h$. 
It is related to the nonlocal susceptibility by $\hat\chi(0) = \lim_{d\to0} \chi(0,x_i,x_j) d$. 

\begin{figure*}
    \centering
    \includegraphics[width=0.7\textwidth]{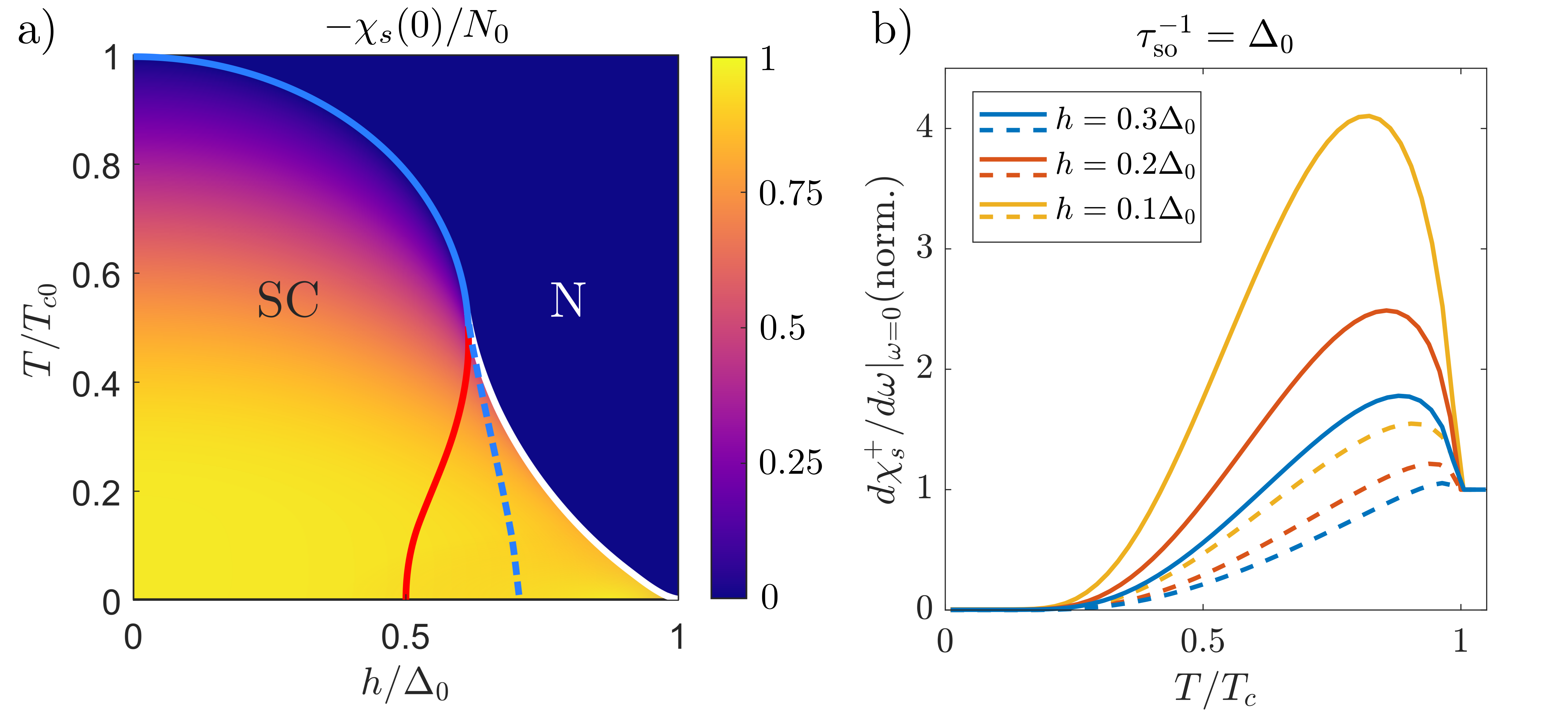}
    \caption{
    a) Thin-film static spin susceptibility in the SC state without spin scattering. Overlaid lines show the phase diagram. Blue solid (dashed) line indicates a second-order (thermodynamic first order) phase transition\cite{maki1964-ppm}. Within the area delimited by red and white curves, both SC and normal state are (meta)stable. 
    b) Dynamic correction for the transverse spin susceptibility. Solid and dashed lines denote the real and imaginary part, respectively. Real and imaginary parts are normalized by their normal state values given by Eq.~\eqref{eqS:normal_spin_susc}.}
    \label{figS:spin_susceptibility}
\end{figure*}

Let us denote by $\chi_{\rm homog}(\omega)$ the usual spin susceptibility to an external in-plane magnetic field \cite{yosida1958-pss}. It is related to the nonlocal thin-film spin susceptibility by a simple shift: $\chi(\omega) = \chi_{\rm homog}(\omega) - N_0$. These susceptibilities vanish at different limits; $\chi_{\rm homog}(0)=0$ at $T=0$ in the absence of spin scattering\,\cite{abrikosov1962-soi}, whereas the nonlocal susceptibility vanishes in the normal state. Figure \ref{figS:spin_susceptibility}a shows the static spin susceptibility for $\tau_{\rm so}^{-1}=\tau_{\rm sf}^{-1}=0$.

The dynamic spin susceptibility $\hat \chi(\omega)$ is defined as a spin response to a dynamic perturbation of the exchange field
\begin{equation}
\bm S(\omega,x_i) = \sum_{j={1,2}}  \hat\chi(\omega,x_i,x_j) G_j \bm m_j(\omega).
\end{equation}
Generally, $\hat\chi$ is a $3\times3$ matrix. In the thin-film limit, it is diagonal in the standard circular basis
\begin{equation}
\begin{split}
\bm e_{\pm} = \mp(\bm x \pm i\bm y)/\sqrt{2},\quad
\bm e_{z} = \bm z,
\end{split}
\end{equation}
where the $z$-axis is chosen along the static component $\bm h(0)$.
The components of $\hat\chi$ are extracted as
$\chi^{\mu\nu} = \bm e_\mu^\dag \hat\chi \bm e_\nu$. In the parallel case, or in the thin-film limit, the susceptibilities $\hat\chi(\omega,x_i,x_j)$ can be simultaneously diagonalized, and we drop the second index: $\chi^+ \equiv \chi^{++}$.
We call $\chi^{z}$ the longitudinal susceptibility and $\chi^{+(-)}$ left(right)-handed transverse susceptibility. The longitudinal component is only induced if $\bm m_{1}(0)$ and $\bm m_{2}(0)$ are non-collinear, since only then $\bm m_1(\omega)\cdot \bm m_2(0) \neq 0$.

In the thin-film limit finite frequency transverse spin susceptibility can be obtained by solving the perturbed Usadel equation 
   \begin{equation}
       \begin{split}
   [\hat\Lambda(1)&+\hat \Sigma(1)] \hat g (12)  - \hat g (12) [\hat \Lambda(2)+\hat \Sigma(2)] \\
    &+ \hat \Sigma(12) \hat g_0(2) - \hat g_0(1) \hat \Sigma(12)\\ 
    &= i [ \hat g_0(1) 
  ({\bm h}(\omega)\cdot\bm \sigma) \tau_3  - 
  ({\bm h}(\omega)\cdot\bm \sigma) \tau_3 
  \hat g_0(2)
  ],
  \end{split}
\end{equation}
where $\hat g_0(1)$ is the static GF at Matsubara frequency $\varepsilon_{1}=i \omega_{n}$ and $\hat g(12)$ is the perturbation induced by the left polarized driving field $\bm h(\omega) = h^+(\omega) \bm e_+$, and
\begin{align}
    \hat\Lambda(\varepsilon) &= -i(\varepsilon - h\sigma_3)\tau_3 + \Delta\tau_1,\\
    \hat s(\varepsilon) &= \sqrt{[\hat \Lambda(\varepsilon)+\hat\Sigma_0(\varepsilon)]^2},\\
    \hat g_0(\varepsilon) &= \hat s(\varepsilon)^{-1}[\hat\Lambda(\varepsilon)+\hat\Sigma_0(\varepsilon)].
\end{align}
Using the normalization condition
\begin{equation}
    \hat g_0(1)^2 = 1,\quad  
    \hat g_0(1) \hat g(12) + \hat g(12) \hat g_0(2) =0,
\end{equation}
and the relation
\begin{equation}
\begin{split}
\tau_3 &\hat g_0(1)\hat g(12)\hat g_0(2)\tau_3 + \hat g_0(1) \tau_3 \hat g(12) \tau_3 \hat g_0(2) \\
&= -2[g_{\uparrow 1}(1)g_{\downarrow 1}(2)-g_{\uparrow 3}(1)g_{\downarrow 3}(2)] \hat g(12),
\end{split}
\end{equation}
the solution for the Matsubara GF is found as
\begin{equation}
    \hat g(12)\\
        = \frac{i [ \tau_3
  - 
  \hat g_\uparrow(1) \tau_3 
  \hat g_\downarrow(2)
  ] \bm h(\omega)\cdot\bm\sigma}{s_\uparrow(1)+s_\downarrow(2) + \frac{1}{3\tau_{\rm so}} + \frac{g_{\uparrow 1}(1)g_{\downarrow 1}(2) - g_{\uparrow 3}(1) g_{\downarrow 3}(2) }{3\tau_{\rm sf}}}.\label{eqS:susceptibility_solution}
\end{equation}
The energy-resolved spin susceptibility is obtained by taking the trace
\begin{equation}
\chi(1,2) = -\frac{N_0}{8}\Tr[(\bm e_+ \cdot\bm \sigma)^\dag\tau_3\frac{\hat g(12)}{h_\omega}] 
\end{equation}

The solution for real frequencies is obtained with analytical continuation
\begin{equation}
\begin{split}
    \chi^{+}(\omega) = \int_{-\infty}^{\infty} d{\varepsilon} \big[ \chi(1^R,2^R) f_2 - f_1 \chi(1^A,2^A)&\\
     + (f_1-f_2) \chi(1^R,2^A)& \big],\label{eqS:analytical_continuation}
\end{split}
\end{equation}
where $1^{R/A}$ and $2^{R/A}$ stand for the replacements $i\omega_n \to \varepsilon \pm i\Gamma$, $i\omega_m \to \varepsilon - \omega \pm i\Gamma$, $f_{1} = \tanh(\varepsilon/(2T))$, $f_2 = \tanh((\varepsilon-\omega)/(2T))$, and $\Gamma$ is an infinitesimally small quantity denoting the correct solution branch. In the numerical solution, we use a finite $\Gamma< 10^{-2}\Delta_0$ to broaden the BCS divergence in the density of states, which makes the numerical solution converge faster.

\begin{figure*}[t]
    \centering
    \includegraphics[width=0.7\textwidth]{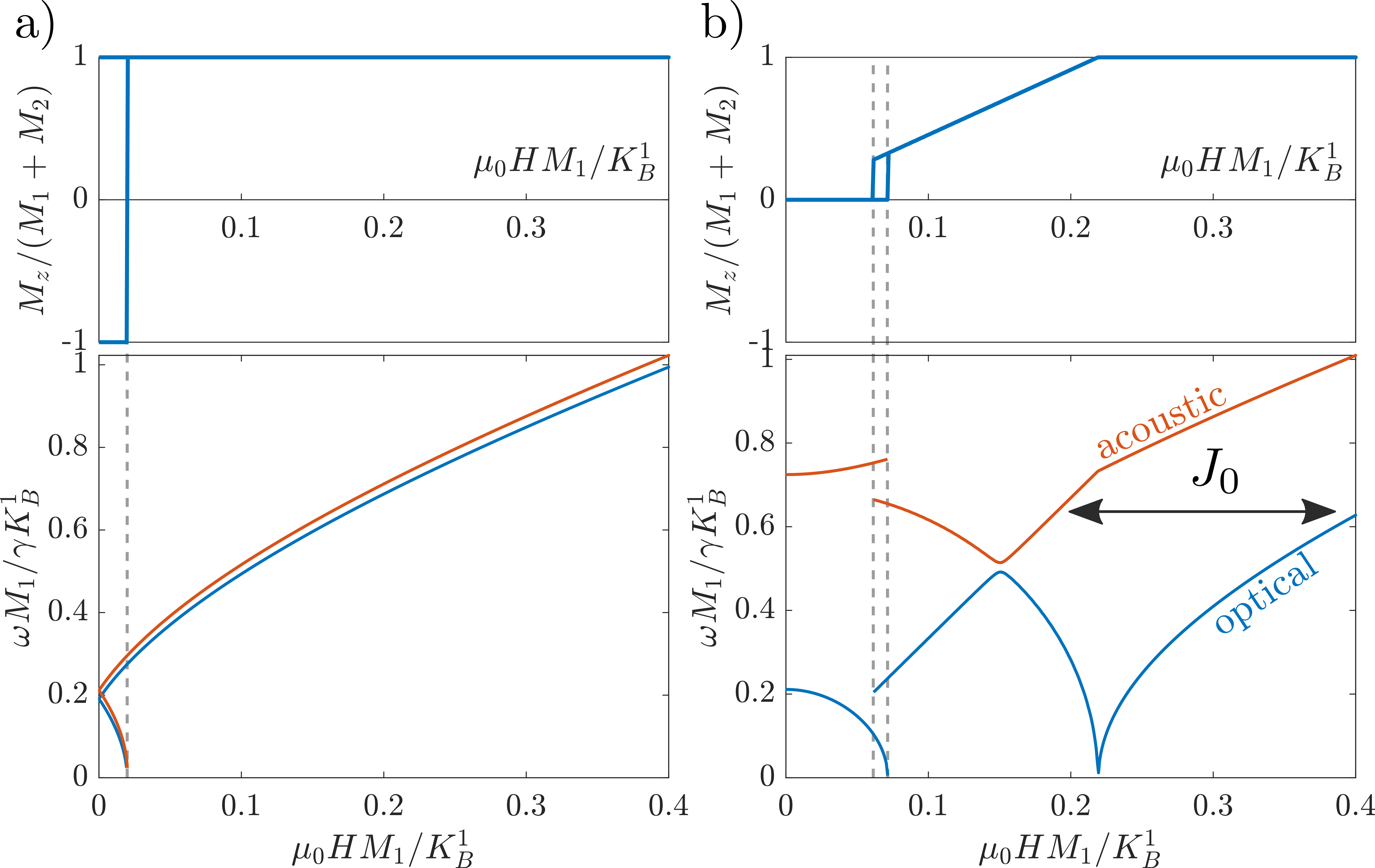}
    \caption{FMR dispersion (bottom) and magnetic hysteresis (top) in a symmetric bilayer a) in the N state and b) in the SC state at $T=0.01T_{c0}$. In the N state the magnets are only weakly coupled by pumped quasiparticle spin currents. In the SC state the magnets are also coupled by SS. SC is evaluated in the thin-film limit. Parameters: $d N_0 \Delta_0^2=0.5$, $K_{A}^{n} = 0.01$, $\Delta_0 M_1/\gamma=10^3$, in units of $K_B^{n}$; $h_{n}=0.3\Delta_0$, $\alpha_{n}=0.01$, $\tau_{\rm so/sf}=0$.  External field is at {1\degree} angle to the anisotropy axis.}
    \label{figS:dispersion}
\end{figure*}

The solution Eq.~\eqref{eqS:susceptibility_solution} depends on the numerically determined equilibrium GFs and thus the integral \eqref{eqS:analytical_continuation} is evaluated numerically. The low-frequency corrections to the static spin susceptibility are shown in Fig.~\ref{figS:spin_susceptibility}b. In the normal state, there are no spectral changes, and only the last term in Eq.~\eqref{eqS:analytical_continuation} contributes:
\begin{equation}
    \chi_{\rm n}^{+}(\omega) = \frac{2N_0 \omega}{\omega - 2h + i\tau^{-1}_{\rm sn}},\label{eqS:normal_spin_susc}
\end{equation}
with $\tau_{sn}^{-1} = 4(\tau_{\rm so}^{-1} + \tau_{\rm sf}^{-1})/3 + 2\Gamma$. We see that $\Gamma$ acts as an additional mechanism for spin relaxation. In general, our results depend on $\Gamma$ only if the dissipation from it is comparable to dissipation from other sources ($\tau_{\rm so/sf}$ and $\alpha_{n}$).

For longitudinal susceptibility, there is no simple analytical solution, but the Usadel equation can be written as a 4-component matrix equation, which is solved numerically\,\cite{ojajarvi2021-ges}. For small frequencies, the longitudinal and transverse susceptibility differ near $T_c$, where a small change in the modulus of $\bm h$ can have a large effect on $\Delta$. 

Since both $\bm h(t)$ and $\bm S(t)$ are real, the left- and right-handed susceptibilities are related by $\chi^{-}(\omega) = \chi^{+}(-\omega)^*$. Also, the transverse spin susceptibility at zero frequency and the static nonlinear spin accumulation can be related to each other by considering an adiabatic rotation of the total exchange field, so that when $\hat \chi(0,x_i,x_j)$ for $i,j\in\{1,2\}$ can be simultaneously diagonalized, we have
\begin{equation}
    \bm S^{\rm eq}(0,x_i) = \sum_{j=1,2} \chi^{\pm}(0,x_i,x_j)G_j\bm m_j(0).\label{eq:static_transverse_connection}
\end{equation}
This relation is needed to show the exact cancellation of spin supercurrents for the acoustic mode in P configuration.

\subsection{Magnetic susceptibility of FI/SC/FI trilayer}

In this section, we derive the non-collinear matrix susceptibility for a coupled magnetic system. To find the susceptibility, one first solves the equilibrium problem by finding a local minimum of the free energy, 
Eq.~(8). 
This determines the static magnetization directions
\begin{equation}
    \bm m_{1/2}(0) \upharpoonleft\!\upharpoonright\bm H^{(0)}_{\rm eff,1/2} - \frac{J_\phi \bm m_{2/1}(0)}{\Ms{1/2}}.
\end{equation}

In LLG equation the modulus of the magnetization is fixed and there are only two degrees of freedom for each magnet. To remove the third dimension, we rotate 
Eq.~(11)
to each magnet's individual eigenbasis in which $\bm m_n(0)$ points in the $z$-direction and project on the $x$-$y$ plane. 

Identifying in 
Eq.~(12)
the parts which are proportional to $\bm M_1(\omega)$ and $\bm M_2(\omega)$, and doing the thin-film approximation $\hat\chi(\omega,x_i,x_j) G_j\approx \hat \chi(\omega)h_j$, we find the magnetic susceptibility for a non-collinear static configuration
 \begin{subequations}
  \begin{align}
     \hat\chi^{-1}_{\rm tot}(\omega) &= \begin{pmatrix}
     \hat\chi^{-1}_1(\omega)+\frac{\hat L_{11}(\omega)}{\Ms{1}^2} & \frac{\hat L_{12}(\omega)}{\Ms{1}\Ms{2}}\\
     \frac{\hat L_{21}(\omega)}{\Ms{1}\Ms{2}} & \hat\chi^{-1}_2(\omega)+ \frac{\hat L_{22}(\omega)}{\Ms{2}^2}
     \end{pmatrix},\label{eqS:magnetic_susceptibility}\\
 \chi^{-1}_i(\omega)^{++} &= \frac{\omega- \gamma \mu_0 H_{\rm eff,i} - i\alpha_i\omega}{\gamma\Ms{i}},\\
  \chi_{i}^{-1}(\omega)^{+-} &= - K_B^i/\Ms{i}^2\\
  \mu_0 H_{\rm eff,i}&= \mu_0 H\cos(\theta-\phi_i)\nonumber\\
  &+ [2K_A^i\cos(\phi_i)^2-K_B^i]/\Ms{i}\\
  &+ {d \chi(0) h_1 h_2 \cos(\phi_1-\phi_2)}/{\Ms{i}}\nonumber,
  \end{align}
where $\phi_i$ ($\theta$) is the angle between the anisotropy axis $\hat{\bm z}$ and $\bm m_i(0)$ ($\bm H$). We assume that the static magnetization is always in-plane.
The spin currents not included into the effective magnetic field are
\begin{equation}
\begin{split}
\hat L_{ij}(\omega) &= -d h_i h_j\left[ \hat D_i\hat\chi(\omega)\hat D_j^\dag\,{-}\,\chi(0)\hat{\openone}_2\delta_{ij}\right]\\
&- d N_0 h_i \omega \delta_{ij} \hat\sigma_3,\label{eqS:dynamic_coupling}
\end{split}
\end{equation}
 \end{subequations}
 where $\hat \chi=\diag(\chi^-,\chi^z,\chi^+)$ is the spin susceptibility in the SC eigenbasis, and the $2\times3$ matrices $\hat D_n=\hat P \hat D^1(0,\phi_n-\phi_{\rm sc},0)$ are a product of a projection $\hat P$ to the $x$-$y$ plane, and of spin-1 Wigner matrices $\hat D^1$. $\phi_{\rm sc}$ is the direction of $\bm h(0)$.

The eigenmodes are found by diagonalizing $\hat\chi^{-1}_{\rm tot}(\omega)$ and finding the values of $\omega$ for which the real part of an eigenvalue vanishes. A typical dispersion for a symmetric system is shown in Fig.~\ref{figS:dispersion}.

In parallel configuration the magnetic susceptibility
can be written as 
Eqs.~(13-14)
with
\begin{subequations}
\begin{align}
    \chi_{i}^{-1}(\omega)^{++} &= \frac{\omega- \gamma \mu_0 H_{\rm eff,i}^{(0)} - i\alpha_i\omega}{\gamma\Ms{i}},\label{eqS:bare_susc_diag2}\\
    \chi_{i}^{-1}(\omega)^{+-} &= - K_B^i/\Ms{i}^2 \label{eqS:bare_susc_off2}\\
    \mu_0 H_{\rm eff,i}^{(0)} &= \mu_0 H + 2K_A^i/\Ms{i} - K_B^i/\Ms{i},
\end{align}
\end{subequations}
and with spin susceptibility $\hat\chi = \diag(\chi^{++},\chi^{--})$ which includes only the transverse components. Here we have separated the N and SC state contributions to the effective fields into $H^{(0)}_{{\rm eff},i}$ and $-J/\Ms{i}$, respectively, and did not assume the thin-film limit.

Since $\bm M_n(t)$ and $\bm H_{\rm rf}(t)$ are real,
the $-+$ and $--$ components can be obtained by $\chi^{-1}(\omega)^{\mu\nu} = [\chi^{-1}(-\omega)^{\nu\mu}]^*$. The spin current $j_{xi}^\mu(\omega) = \xi_{ij}^\mu(\omega) m_j^\mu(\omega)$, where $\xi$ is a response function, also obeys the same reality condition $\xi_{ij}^-(\omega) = [\xi_{ij}^+(-\omega)]^*$.

\subsection{FMR linewidth}

Here we give the definition of the FMR linewidth used in the main text.
Typically in FMR experiment, rf frequency is held fixed while sweeping the external field strength $H$.  The power of the rf drive is given by \cite{tserkovnyak2005-nmd}
\begin{equation}
\begin{split}
    P/A &= \sum_{n=1,2} \langle \bm H_{\rm rf}(t)\,{\cdot}\,\partial_t{\bm M}_n(t)\rangle\\
    &= \sum_{\lambda\in{\text{eigs}}} \omega \Im \chi_\lambda(\omega) |H_\lambda|^2,
\end{split}
\end{equation}
where $\chi_\lambda$ are the eigenvalues of the matrix susceptibility $\hat\chi$, and $H_\lambda$ are the projections of the rf field along the corresponding eigenvector.

The linewidth is defined as the difference $\Delta H = (2/\sqrt 3)(H_{\rm min}-H_{\rm max})$ between the minimum and the maximum of the field derivative $d P/dH$. 
Near resonance, $dP/dH$ is dominated by the resonant eigenmode $\lambda$ and other modes can be neglected. To determine the linewidth, we linearize the susceptibility of the resonant eigenmode, expressing it in terms of resonance field $\mu_0 H_{\rm res}$, weight factor $W$ and linewidth $\mu_0\Delta H$, 
\begin{equation}
\chi_{\lambda}(H)^{-1} \approx \frac{\mu_0(H - H_{\rm res}) - i\mu_0\Delta H
}{W}.
\end{equation}
The above parameters are defined by the equations
\begin{subequations}
\begin{align}
    0 &= \Re[\chi_{\lambda}(H_{\rm res})^{-1}],\\
    \frac 1 {W} &= \left.\frac{{\rm d}\Re[\chi_{\lambda}(\omega)^{-1}]}{{\rm d}\mu_0H}\right|_{H=H_{\rm res}}, \\
    \mu_0\Delta H &= -W\Im[\chi_{\lambda}(H_{\rm res})^{-1}].\label{eqS:aeffdef}
\end{align}
\end{subequations}
In bulk ferromagnets, the field linewidth can be identified with Gilbert damping $\alpha = \gamma\mu_0\Delta H/\omega$.

\subsection{Effect of the first order SC transition on the magnetic hysteresis}

\begin{figure*}
    \centering
    \includegraphics[width=0.9\textwidth]{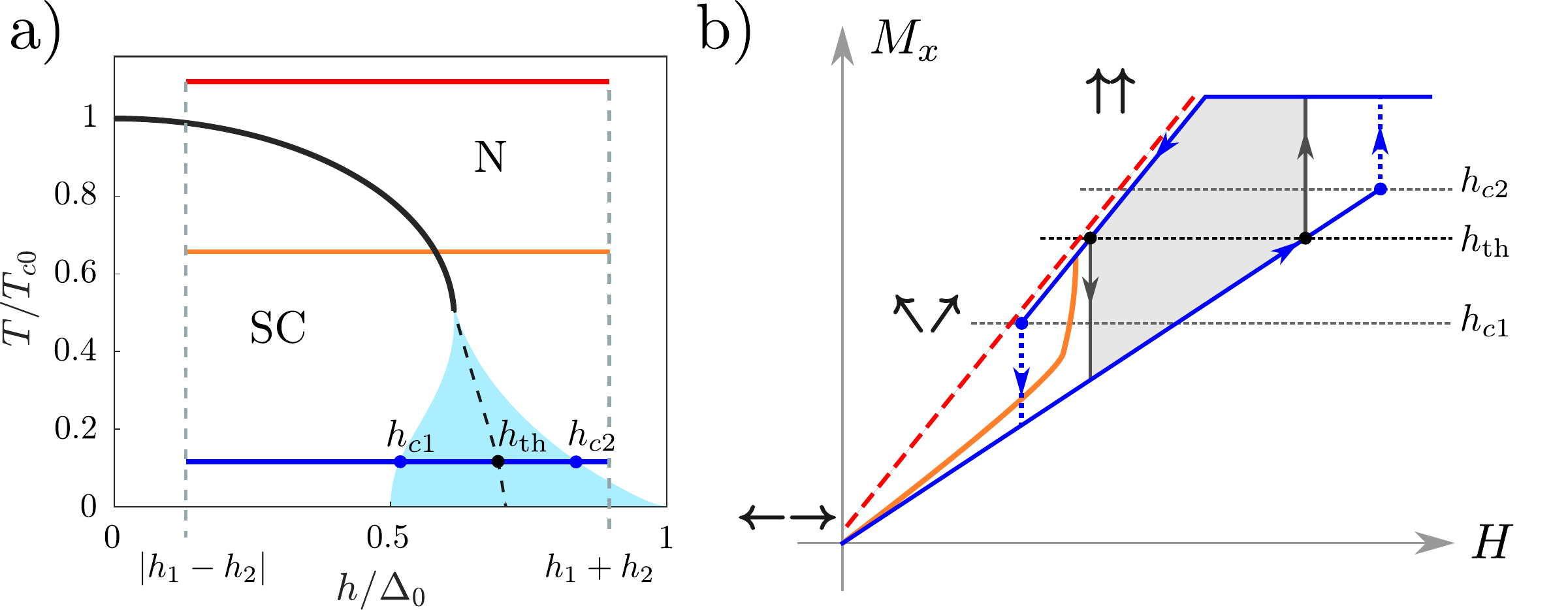}
    \caption{Effect of the first order SC transition on the magnetic hysteresis. a) Phase diagram as a function of temperature and exchange field. Red, orange and blue lines show the range of total exchange fields, when the individual magnets induce a field of strength $h_1$ and $h_2$. Blue shading indicated the region of metastability where both SC and N state are possible. Dashed black line indicates the thermodynamic phase transition. b) Magnetization along the external field ($M_x$) as a function of the field strength $H$. Red dashed line and the solid orange and blue lines correspond to the temperatures indicated in panel (a). Red dashed line is the high temperature (normal state) behavior. Orange line is the behavior at intermediate temperatures when the SC transition is continuous. Blue line shows the behavior at low temperatures in which the SC transition is of the first order.
    }
    \label{fig:normalSC-hysteresis}
\end{figure*}

At low temperatures, the transition between SC and the N state as a function of the exchange field is of the first order.\cite{maki1964-ppm,tedrow1970experimental} In the phase diagram, the first order transition is accompanied by a region in which both phases are possible as metastable states, albeit only one of them is thermodynamically stable away from the phase boundary (Fig.~\ref{fig:normalSC-hysteresis}a). Experimentally, the metastability shows as hysteresis or state memory when the induced exchange field is swept over the phase boundary.\cite{wu1995asr,butko1999state}

The FI/SC/FI trilayer provides a way of probing the transition and the SC-N hysteresis via magnetization measurements. The requirement for such measurement is that it must be possible to vary the total exchange field induced into the SC continuously as a function of the external field. This requirement is met if the FIs have an in-plane easy-axis and the magnetic field is applied perpendicular to it. 

The FIs individually induce the exchange fields $h_1, h_2 >0$ into the SC. In the AP and P configurations, the total fields are $|h_1-h_2|$ and $h_1+h_2$, respectively. By changing the magnetization configuration, we thus sweep a horizontal segment of the phase diagram as shown in Fig.~\ref{fig:normalSC-hysteresis} for three different temperatures (blue, orange and red lines). If $h_1$ and $h_2$ are chosen properly, we cross the first order phase boundary at a low temperature (blue). Let us denote the thermodynamic transition point by $h_{\rm th}$. If there is a measurable SC-N hysteresis, the N-to-SC and SC-to-N transitions occur at fields $h_{c1}<h_{\rm th}$ and $h_{c2}>h_{\rm th}$, respectively. 

Let us then consider the magnetization in a symmetric trilayer as the external field is varied. In the normal state the magnets are uncoupled. At zero field the magnetizations lie along the anisotropy axis. Increasing the field twists the magnetizations until they point towards the external field (red dashed line in Fig.~\ref{fig:normalSC-hysteresis}). The P and AP configurations are degenerate and when the field is reduced to zero, the system may end up in either one. 

The SC state on the other hand favors the AP configuration, and the slope of the magnetization curve (orange and blue lines) is lower. The first order transition (blue line) shows in the magnetization curve as hysteresis. However, the magnetic hysteresis itself is not necessarily evidence of SC-N hysteresis, but only indicates a first-order transition between SC and N states. The shaded region in Fig.~\ref{fig:normalSC-hysteresis} is the magnetic hysteresis obtained by assuming that the SC transition always occurs when $h$ crosses the thermodynamic field $h_{\rm th}$, with no hysteresis. With this assumption, the P-to-AP and AP-to-P transitions occur in Fig.~\ref{fig:normalSC-hysteresis}b at the same total magnetization, but the field strength depends on the direction of transition.

The evidence for SC-N hysteresis is the difference in the magnetization $M$ at which the transitions occur, since there is a one-to-one correspondence between $M$ and $h$.

%



\end{document}